\documentclass[twocolumn,showpacs,amsmath,amssymb,superscriptadress,prc]{revtex4-1}
\usepackage{epsfig,amsmath,graphicx,amssymb,tabularx}
\usepackage[dvipdfmx,bookmarks=true,colorlinks,citecolor=blue,linkcolor=blue,anchorcolor=blue,filecolor=blue,urlcolor=blue]{hyperref}
\usepackage[T1]{fontenc}
\usepackage{CJK}
\usepackage{dcolumn}
\usepackage{bm}
\usepackage{multirow}

\begin{document}
\begin{CJK*}{GBK}{}

\title{Isotopic trends of quasifission and fusion-fission in the reactions $^{48}$Ca+$^{239,244}$Pu}

\author{Lu Guo$^{1,2}$}
\email{luguo@ucas.ac.cn}
\author{Caiwan Shen$^{3}$}
\author{Chong Yu$^{4}$}
\author{Zhenji Wu$^{4}$}

\affiliation{$^{1}$School of Nuclear Science and Technology, University of Chinese Academy of Sciences, Beijing 100049, China}
\affiliation{$^{2}$Institute of Theoretical Physics, Chinese Academy of Sciences, Beijing 100190, China}
\affiliation{$^{3}$School of Science, Huzhou University, Huzhou 313000, China}
\affiliation{$^{4}$School of Physics, University of Chinese Academy of Sciences, Beijing 100049, China}

\date{\today}

\begin{abstract}
\textbf{Background:} Quasifission and fusion-fission are primary
mechanisms to prevent the production of superheavy elements. The recent
experimental measurements reveal that the fusion-evaporation cross
section in the $3n$ reaction channel of $^{48}$Ca+$^{239}$Pu is 50 times
lower than using $^{244}$Pu as target nucleus
[\href{https://journals.aps.org/prc/abstract/10.1103/PhysRevC.98.014615}{Phys. Rev. C 92, 034609 (2015)}].
However, the precise mechanisms of this remarkable isotopic
dependence are not well understood.

\textbf{Purpose:} To understand the experimental observation of the
rapid decrease of stability of superheavy nuclei as the neutron number decreases,
the theoretical studies of quasifission and fusion-fission in connection
with experimental production for $Z$=114 flerovium isotopes are
required to investigate the possible differences in reaction mechanisms induced
by these two targets.

\textbf{Methods:} We propose an approach called TDHF+HIVAP to take into account both the evolution of dinuclear system
and the deexcitation of compound nucleus, which combines the microscopic time-dependent
Hartree-Fock (TDHF) method for the fusion and quasifission
dynamics with the statistical evaporation model HIVAP for fusion-fission dynamics.

\textbf{Results:} Fusion is observed for both reactions $^{48}$Ca+$^{239,244}$Pu with the side orientation the deformed target nucleus,
while quasifission dynamics is observed for the tip orientation. The
nuclear contact times, masses and charges as well as the kinetic energies of
the fragments, and the mass-angle distribution strongly depend on the
colliding energy, impact parameter, and deformation orientation.
The quantum shell effect displays a crucial role in both the quasifission and the fusion-fission processes.
The quasifission is considerably reduced and the survival probability is enhanced around one order of magnitude
in the reaction using $^{244}$Pu target as compared to the $^{239}$Pu case.

\textbf{Conclusions:}  The studies by using TDHF+HIVAP method well
account for the experimental observations and the present method clearly
shows its applicability in the reaction mechanisms of quasifission and fusion-fission dynamics.
The experimental and theoretical results encourage the use of neutron-rich targets for the
production of new superheavy elements.

\end{abstract}
\maketitle

\end{CJK*}

\section{Introduction}

\label{introduction}

The creation of superheavy elements (SHEs) is one of the most challenging
research topics in nuclear physics. SHEs up to proton number $Z=118$ have been
experimentally produced in fusion-evaporation reactions, either using $^{208}%
$Pb (or $^{209}$Bi) as the target in cold fusion, or $^{48}$Ca-induced hot
fusion colliding with the actinide nuclei. However, the SHEs produced so far
in the experiments are far from the long-lived stability island, which was
predicted by many theoretical
approaches~\cite{Bender2001_PLB515-42,Cwiok2005_Nature433-705,Pei2005_PRC71-034302}%
, locating at the neutron magic number $N=184$ and proton magic number
$Z=114-126$ as a result of new shell closures. The existence of the stability
island is supported by the observed increase of stability of heavier isotopes
approaching to the predicted magic number $N=184$%
~\cite{Oganessian2010_PRL104-142502,Ellison2010_PRL105-182701,Kozulin2014_PRC90-054608}.

The entrance channel dynamics is critical for SHE formation. In heavy-ion
collisions, capture reaction produces a dinuclear system. The shape evolution
of this initial fragile dinucleus (either fusion or quasifission) is
determined by the dynamical dissipation from the collective kinetic energy to
internal degrees of freedom. The dinucleus system may evolve to an
equilibrated compound nucleus in fusion process. Alternatively it can also
break apart, characterized by the massive nucleon transfer and contact time
longer than 5 zs. This process is known as quasifission (QF)
~\cite{Bock1982_NPA388-334,Toke1985_NPA440-327,Shen1987_PRC36-115}. For light
and medium-mass systems typically with $Z_{p}Z_{t}<1600$, the compound nucleus
will be formed once the projectile is captured by the target. For heavy
systems producing SHE, however, the fusion probability is dramatically reduced
by the quasifission. Furthermore, even if the compound nucleus survives
against quasifission, the fusion-fission (FF) has very large probability to
happen due to its excitation. Hence, the production cross section forming SHE
is substantially reduced by the quasifission and fusion-fission processes.

To produce the new SHEs and heavier isotopes of known SHE experimentally, the
optimal target-projectile combination and bombarding energy should be chosen
to have the highest residue cross sections for the desired SHE. Experimental
measurements indicate an increase of residue cross sections for the reactions
involving more neutron-rich
nuclei~\cite{Dragojevic2008_PRC78-024605,Oganessian2013_PRC87-034605,Utyonkov2015_PRC92-034609}.
From the theoretical point of view,
the enhancement of residue cross section in the reactions with neutron-rich
target could arise from the decrease of QF and FF probability. However, the
precise mechanisms, especially for the interplay between the QF and FF process, are not well
understood. The dependence of QF on the neutron richness
of compound nucleus is supported by the recent experimental measurements of
mass-angle distributions~\cite{Itkis2011_PRC83-064613,Rietz2011_PRL106-052701,Lin2012_PRC85-014611,Wakhle2014_PRL113-182502}.
In particular, the recent studies of QF show a great promise for a deep insight of
reaction mechanism. Meanwhile QF dynamics may be affected by many
variables, e.g., collision
energy~\cite{Hinde2008_PRL100-202701,Nishio2012_PRC86-034608}, deformation and
orientation~\cite{Hinde1996_PRC53-1290,Prasa2016_PRC93-024607}
, shell structure~\cite{Kozulin2014_PRC89-014614,Prasad2015_PRC91-064605}
of the colliding nuclei, and neutron richness of compound
nuclei~\cite{Hammerton2015_PRC91-041602}. Besides the competition between fusion and quasifission, the competition between
the neutron emission and fission of the excited compound nucleus is also crucial for the
synthesis of SHEs. To understand the complex interplay
of these dynamical processes, it is required to carry out the theoretical studies.

Various theoretical models, including both the macroscopic
models~\cite{Shen2002_PRC66-061602,Feng2009_PRC80-057601,Shen2011_PRC83-054620,
Shen2014_SciChinaPMA57-453,Zhu2014_PRC90-014612,
Zhu2016_PRC93-064610,Wang2016_SciChinaPMA59-642002,Wang2017_ADNDT114-281} and
the microscopic
approaches~\cite{Simenel2012_EPJA48-152,Nakatsukasa2016_RMP88-045004,
Wang2002_PRC65-064608,Wen2013_PRL111-012501,Wen2014_PRC90-054613}%
, have been developed to account for the experimental observations. Although
the macroscopic model may well reproduce the experimental data, a need for
external parameters and a lack of dynamical effect restrict its predictive
power for reactions where no experimental data are available. In present
study, we combine the microscopic time-dependent Hartree-Fock (TDHF) approach with the statistical evaporation model HIVAP
to investigate the effect of neutron richness on the QF and FF dynamics.
TDHF approach provides a profound understanding of nuclear dynamics, as seen from the recent
applications in fusion~\cite{Simenel2004_PRL93-102701,Umar2006_PRC73-054607, Guo2012_EPJWoC38-09003,Umar2012_PRC85-055801,Simenel2013_PRC88-064604,Umar2014_PRC89-034611,
Washiyama2015_PRC91-064607,Godbey2017_PRC95-011601,Simenel2017_PRC95-031601,Schuetrumpf2017_PRC96-064608,Guo2018_PLB782-401},
quasifission~\cite{Golabek2009_PRL103-042701,Oberacker2014_PRC90-054605,Umar2015_PRC92-024621,Umar2016_PRC94-024605,Yu2017_SciChinaPMA60-092011}, transfer reaction~\cite{Washiyama2009_PRC80-031602,Simenel2010_PRL105-192701,
Scamps2013_PRC87-014605,Sekizawa2013_PRC88-014614,Wang2016_PLB760-236,Sekizawa2016_PRC93-054616,Scamps2017_PRC95-024613,Sekizawa2017_PRC96-014615},
fission~\cite{Simenel2014_PRC89-031601,Scamps2015_PRC92-011602,Goddard2015_PRC92-054610,Goddard2016_PRC93-014620,Bulgac2016_PRL116-122504,Tanimura2017_PRL118-152501},
deep inelastic collisions ~\cite{Maruhn2006_PRC74-027601,Guo2007_PRC76-014601,Guo2008_PRC77-041301,Dai2014_PRC90-044609,Dai2014_SciChinaPMA57-1618,
Stevenson2016_PRC93-054617,Guo2017_EPJWoC163-00021,Shi2017_NPR34-41,Umar2017_PRC96-024625},
and resonances dynamics ~\cite{Maruhn2005_PRC71-064328,Nakatsukasa2005_PRC71-024301,Umar2005_PRC71-034314,Reinhard2007_EPJA32-19,Simenel2009_PRC80-064309,
Fracasso2012_PRC86-044303}.
The statistical evaporation model HIVAP~\cite{Shen2002_PRC66-061602,Shen2008_IJMPE17-66} is adopted to take into account the deexcitation process
including both fission and particle evaporation. We will show that the fission barrier of compound nucleus, dominated mainly by the quantum shell effect,
plays significant effect on the fusion-fission process in the present systems.

This article is organized as follows. In Sec.~\ref{theory} TDHF approach with
Skyrme energy functional and fusion-evaporation dynamics are briefly recalled. Section~\ref{discuss} presents
the theoretical analysis of the influence of neutron rich target in the
QF and FF dynamics for the reactions $^{48}$Ca+$^{239,244}%
$Pu. A summary is given in Sec.~\ref{summary}.

\section{Theoretical framework}

\label{theory} In TDHF approach the many-body wave function $\Psi(\mathbf{r},
t)$ is approximated as a single Slater determinant composed by the single-particle states $\phi_{\mathrm{\lambda}}(\mathbf{r}, t)$
\begin{equation}
\Psi(\mathbf{r}, t)=\frac{1}{\sqrt{N!}}\text{det}\{\phi_{\mathrm{\lambda}%
}(\mathbf{r}, t)\},
\end{equation}
and this form is kept at all times in the dynamical evolution. This
approximation leads to the omission of two-body nucleon-nucleon correlations.
By taking the variation of time-dependent action
\begin{equation}
S=\int_{t_{1}}^{t_{2}} dt \langle\Psi(\mathbf{r}, t)|H-i\hbar\partial_{t}%
|\Psi(\mathbf{r}, t)\rangle
\end{equation}
with respect to the single-particle states, one may obtain a set of nonlinear
coupled TDHF equations in the multidimensional spacetime phase space
\begin{equation}
i\hbar\frac{\partial}{\partial_{t}}\phi_{\mathrm{\lambda}}(\mathbf{r},
t)=h\phi_{\mathrm{\lambda}}(\mathbf{r}, t),
\end{equation}
where $h$ is the HF single-particle Hamiltonian. It describes the time
evolution of the single-particle wave functions in a mean field. The set of nonlinear TDHF
equations have been solved on three-dimensional coordinate space without any symmetry
restrictions and with much more accurate numerical methods.

Most TDHF calculations have been done with Skyrme effective
interaction~\cite{Skyrme1956_PM1-1043}. It is natural to represent the Skyrme
force with the energy density functional (EDF), in which the total energy of
the system
\begin{equation}
E=\int d^{3}r \mathcal{H}(\rho, \tau, {\mathbf{j}}, {\mathbf{s}}, {\mathbf{T}%
}, J; {\mathbf{r}})
\end{equation}
is expressed as an integral of the energy functional. The number density
$\rho$, kinetic density $\tau$, current density ${\mathbf{j}}$, spin density
${\mathbf{s}}$, spin-kinetic density ${\mathbf{T}}$, and spin-current
pseudotensor density $J$ are obtained as a sum over single-particle wave
functions. The Skyrme EDF is then expressed as
\begin{align}
\label{EDFH}
\begin{split}
\mathcal{H}  &  =\mathcal{H}_{0}+\sum_{\mathrm{t=0,1}}\Bigg\{A_{\mathrm{t}
}^{\mathrm{s}}\mathbf{s}_{\mathrm{t}}^{2}+A_{\mathrm{t}}^{\Delta{s}}
\mathbf{s}_{\mathrm{t}}\cdot\Delta\mathbf{s}_{\mathrm{t}}\\
&  +A_{\mathrm{t}}^{\mathrm{T}}\bigg(\mathbf{s}_{\mathrm{t}}\cdot
\mathbf{T}_{\mathrm{t}}- \sum_{\mu,\nu=x}^{z}J_{\mathrm{{t},\mu\nu}
}J_{\mathrm{{t},\mu\nu}}\bigg)\Bigg\},
\end{split}
\end{align}
where $\mathcal{H}_{0}$ is the simplified Skyrme functional used in Sky3D
code~\cite{Maruhn2014_CPC185-2195} and most TDHF calculations. For the
definition of coupling constants $A$, see Ref.~\cite{Lesinski2007_PRC76-014312}. Due to the
computational complexity, various approximations to Skyrme EDF have been
employed in TDHF calculations, which restrict the number of degrees of freedom
accessible during a collision, and hence the nature and degree of dissipation
dynamics. For instance, the inclusion of spin-orbit
interaction~\cite{Umar1986_PRL56-2793} solved an early conflict between TDHF
predictions and experimental observation, and turned out to play an important role in
fusion and dissipation
dynamics~\cite{Maruhn2006_PRC74-027601,Dai2014_PRC90-044609}. The time-odd
terms appearing in Eq.~(\ref{EDFH}) are also shown to be non-negligible in heavy-ion
collisions~\cite{Umar2006_PRC73-054607}. Our HF and TDHF codes contain all of
the time-even and time-odd terms in the energy functional and don't impose
the time-reversal invariance, which allow us to compute directly the odd system, such as
those studied here, without resorting to the filling approximation. As pointed
out in Refs.~\cite{Lesinski2007_PRC76-014312,Stevenson2016_PRC93-054617}, the
terms containing the gradient of spin density may cause the spin instability
both in nuclear structure and reaction studies, so we set $A_{\mathrm{t}%
}^{\Delta{\mathbf{s}}}=0$ in our calculations.

In fusion-evaporation reaction, the production cross section for the
superheavy evaporation residues (ER) can be defined as
\begin{equation}
\label{ER}
\sigma_{\rm ER}=\sum_{J=0}^{\infty}\sigma_{\rm cap}
(E_{\rm c.m.},J)P_{\rm fus}(E^*,J)W_{\rm sur}(E^*,J),
\end{equation}
where $\sigma_{\rm cap}$ is the capture cross section for the
projectile and target to come together, $P_{\rm fus}$ is the fusion probability of
dinuclear system against QF, and $W_{\rm sur}$ is the survival
probability of compound nucleus against FF.
Fusion occurs when the collective kinetic energy is entirely
converted into the internal excitation of a well-defined compound nucleus.
Traditionally, the fusion cross section is given by
\begin{equation}
\sigma_{\rm fus}(E_{\rm c.m.})=\sum_{J=0}^{\infty}\sigma_{\rm cap}
(E_{\rm c.m.},J)P_{\rm fus}(E^*,J).
\end{equation}
Since TDHF theory describes the collective motion of fusion dynamics
in terms of semi-classical trajectories, the sub-barrier tunneling of the many-body
wave function can't be included in TDHF. Consequently, the fusion
cross section can be estimated by the quantum sharp-cutoff formula ~\cite{Bonche1978_PRC17-1700}
\begin{equation}
\label{FCS}
\sigma_{\rm fus}(E_{\rm c.m.})=\frac{\pi\hbar^2}{2
\mu E_{\rm c.m.}}[(l_{\rm max}+1)^2-(l_{\rm min}+1)^2],
\end{equation}
where $\mu$ is the reduced mass of the system and $E_{\rm c.m.}$
the initial center-of-mass (c.m.) energy. The quantities $l_{\rm max}$ and $l_{\rm min}$ denote
the maximum and minimum orbital angular momentum for which fusion happens.

In the production of superheavy elements, the excited compound nucleus may undergo the deexcitation process,
which is dominated by fission barrier and neutron evaporation. The survival probability of compound nucleus $W_{\rm sur}$ in Eq.~(\ref{ER})
for $x$ neutron emissions is written as
\begin{equation}
W_{\rm sur}(xn)=\prod_{i=1}^{x}P_{in},
\end{equation}
where $P_{in}$ is the probability of $i$-th neutron emission and given by
the statistical evaporation model~\cite{Weisskopf1937_PR52-295,Bohr1939_PR56-426}
\begin{equation}
P_{in} = \frac{\Gamma_{in}}{\Gamma_{in}+\Gamma_{if}},
\end{equation}
with $\Gamma_{in}$ the $i$-th neutron evaporation width and $\Gamma_{if}$ the $i$-th fission width.
In the numerical calculations of present work,
the \textsc{hivap} code~\cite{Shen2002_PRC66-061602,Shen2008_IJMPE17-66} is adopted to calculate the survival probability $W_{\rm sur}$.

\section{results}

\label{discuss}

TDHF approach has recently demonstrated its feasibility and success in fusion
and quasifission dynamics~\cite{Oberacker2014_PRC90-054605,Umar2015_PRC92-024621,Umar2016_PRC94-024605},
in which the theoretical investigations on the reaction mechanism induced by
the different projectiles colliding the actinide target nucleus account for the
experimental observations reasonably. In present work, our goal is to
investigate how the neutron richness in target nucleus affects the QF and FF dynamics
in the production of SHE flerovium with the reactions $^{48}%
$Ca+$^{239,244}$Pu. We employ the Skyrme SLy5
force~\cite{Chabanat1998_NPA635-231,Chabanat1998_NPA643-441} including all of
the time-odd terms in the mean-field Hamiltonian in our TDHF calculations.

In the numerical simulation, first we calculate an accurate static ground
state for the projectile and target nucleus on the symmetry-unrestricted three-dimensional
grid. The numerical coordinate boxes for the static HF wave functions are
chosen as $24\times24\times24$ fm$^{3}$ for $^{48}$Ca and $32\times28\times32$
fm$^{3}$ for $^{239,244}$Pu, respectively. The nucleus $^{48}$Ca shows a
spherical ground state and $^{239,244}$Pu exhibits a prolate quadrupole
deformation, which are in agreement with experimental data and other calculations. The correct
description of the initial shape of target and projectile nucleus is important
for the dynamical evolution of heavy-ion collisions. Second, we apply a boost
operator on the static single-particle wave functions. The nucleus is assumed
to move on a pure Coulomb trajectory until the initial distance so that the
initial boost is properly treated in TDHF evolution. The time propagation is
performed using a Taylor-series expansion up to the sixth order of the unitary
mean-field propagator and a time step of $0.2~\mathrm{fm}/c$. For the TDHF
dynamical evolution, we use a numerical box of $60\times28\times46$ fm$^{3}$
and a grid spacing of 1.0~fm. The reaction is taken in $x$-$z$ plane and along
the collision axis $x$. The initial separation
between the two nuclei is set to be 30~fm. The choice of these parameters assures a good
numerical accuracy for all the cases studied here. The total TDHF energy and
particle number are well conserved and shift less than 0.1~MeV and 0.01 in the
dynamical evolution, respectively.

\begin{figure}[tb]
\centering
\includegraphics[scale=0.12]{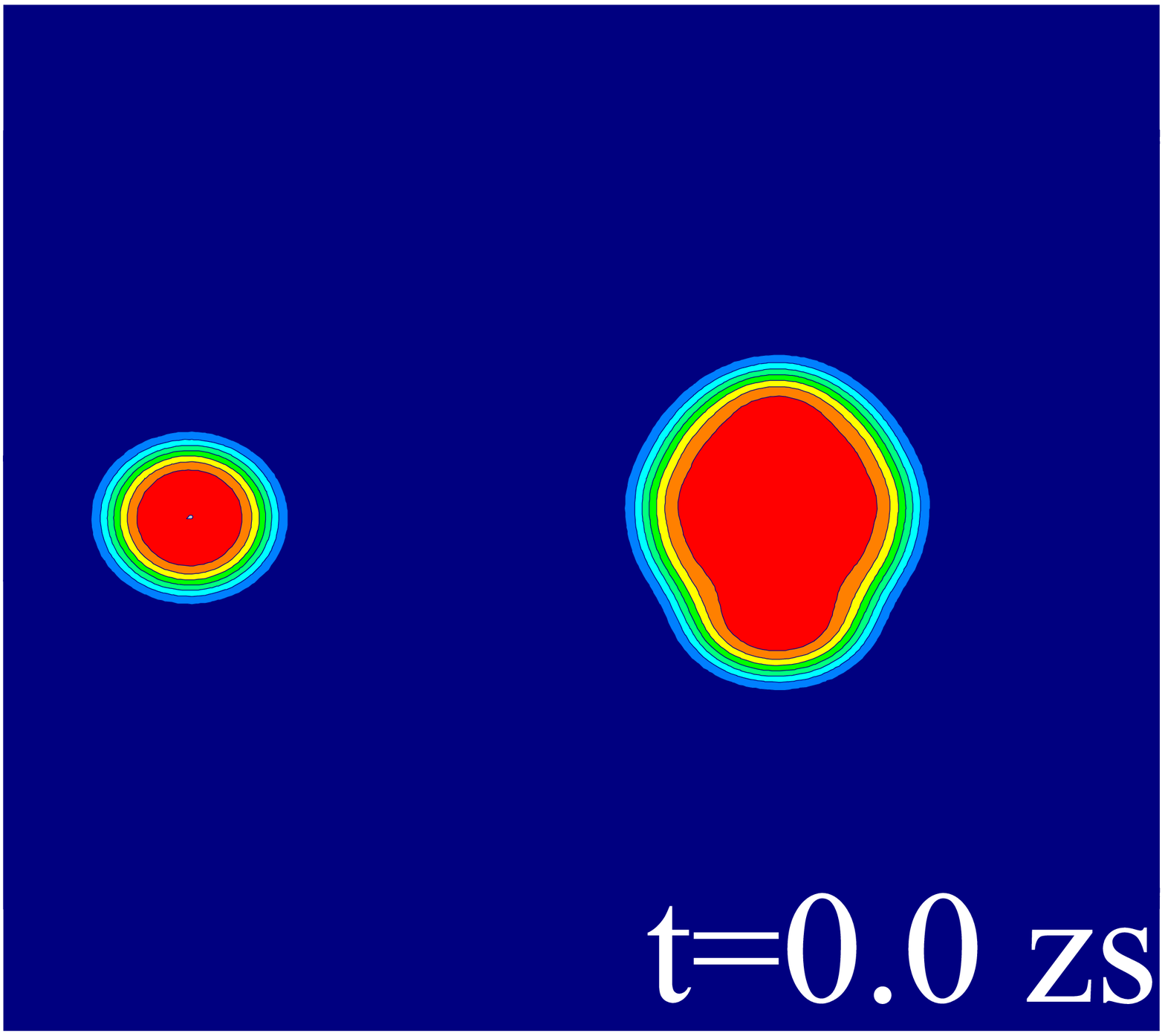}\hspace {1.2pt}
\includegraphics[scale=0.12]{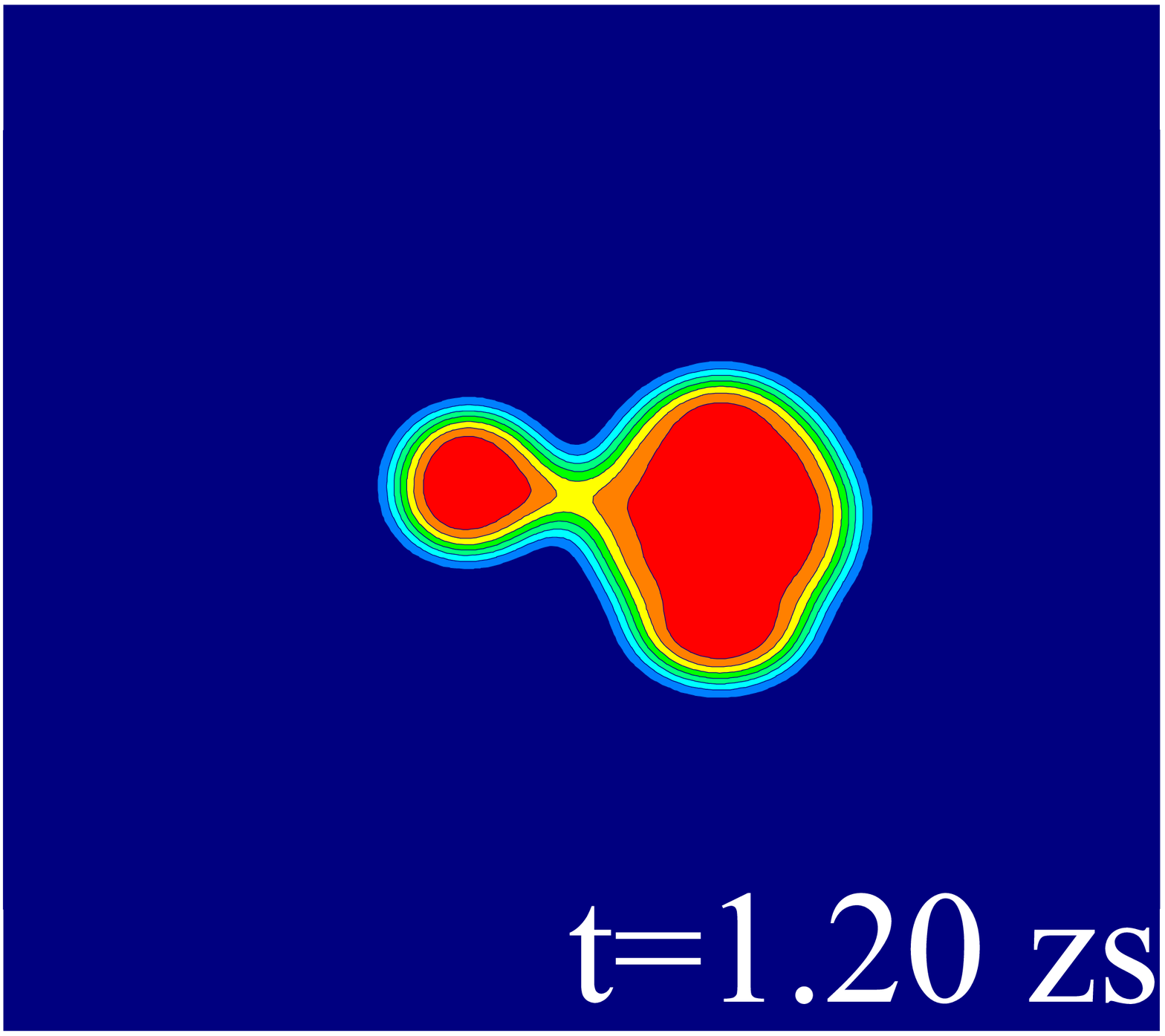}\hspace {1.2pt}
\includegraphics[scale=0.12]{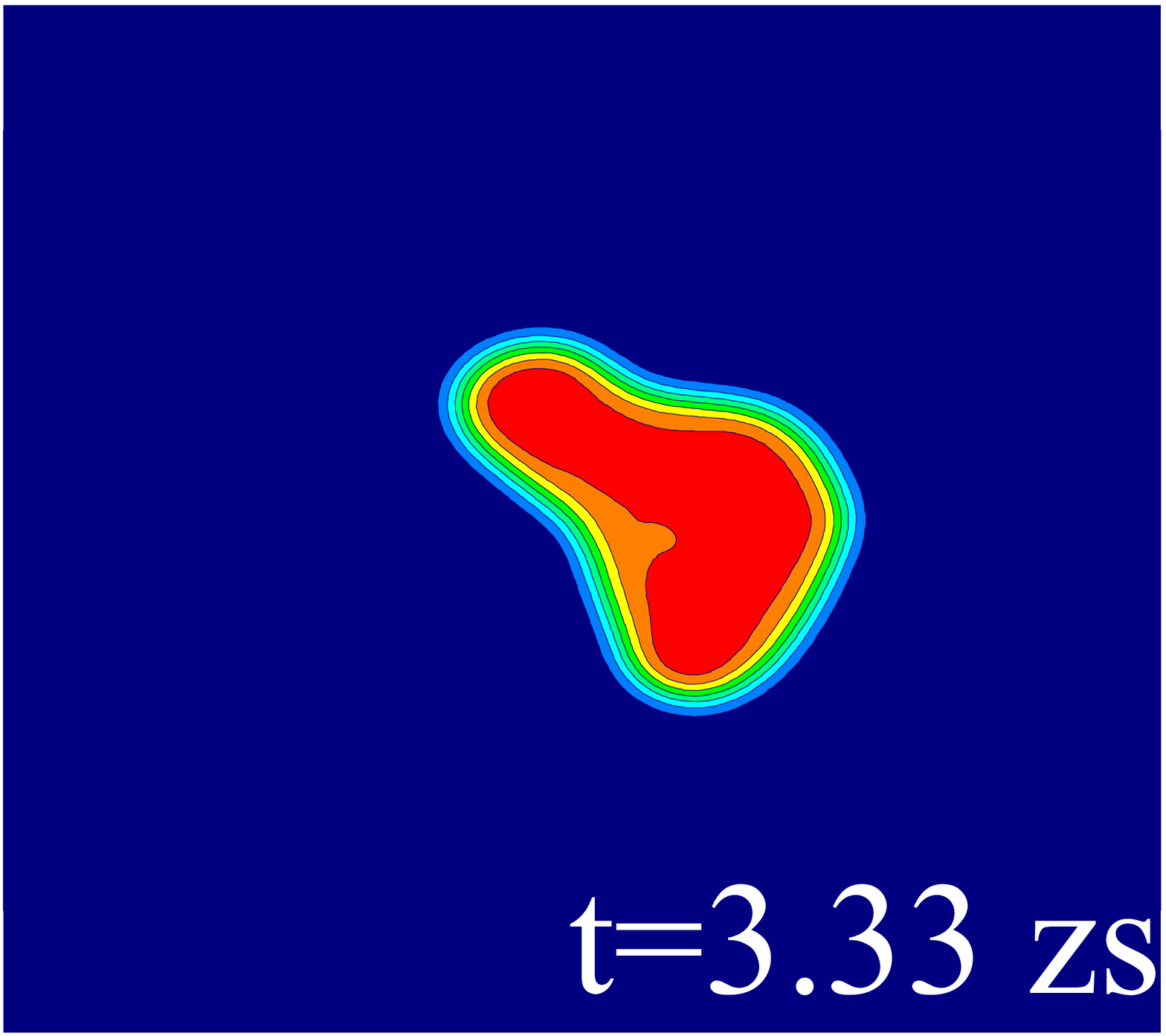}\\[-0.5pt]
\includegraphics[scale=0.12]{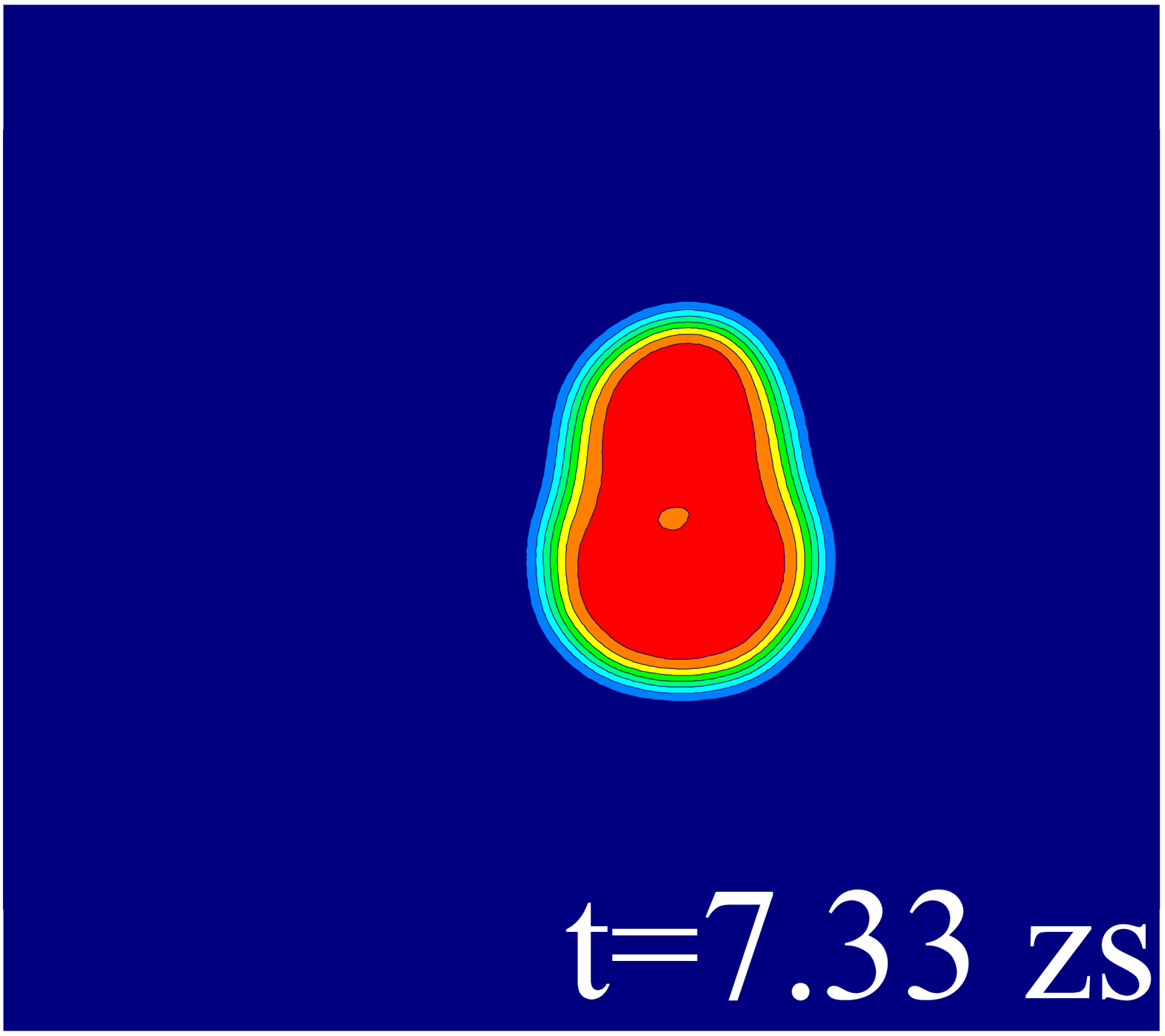}\hspace {1.2pt}
\includegraphics[scale=0.12]{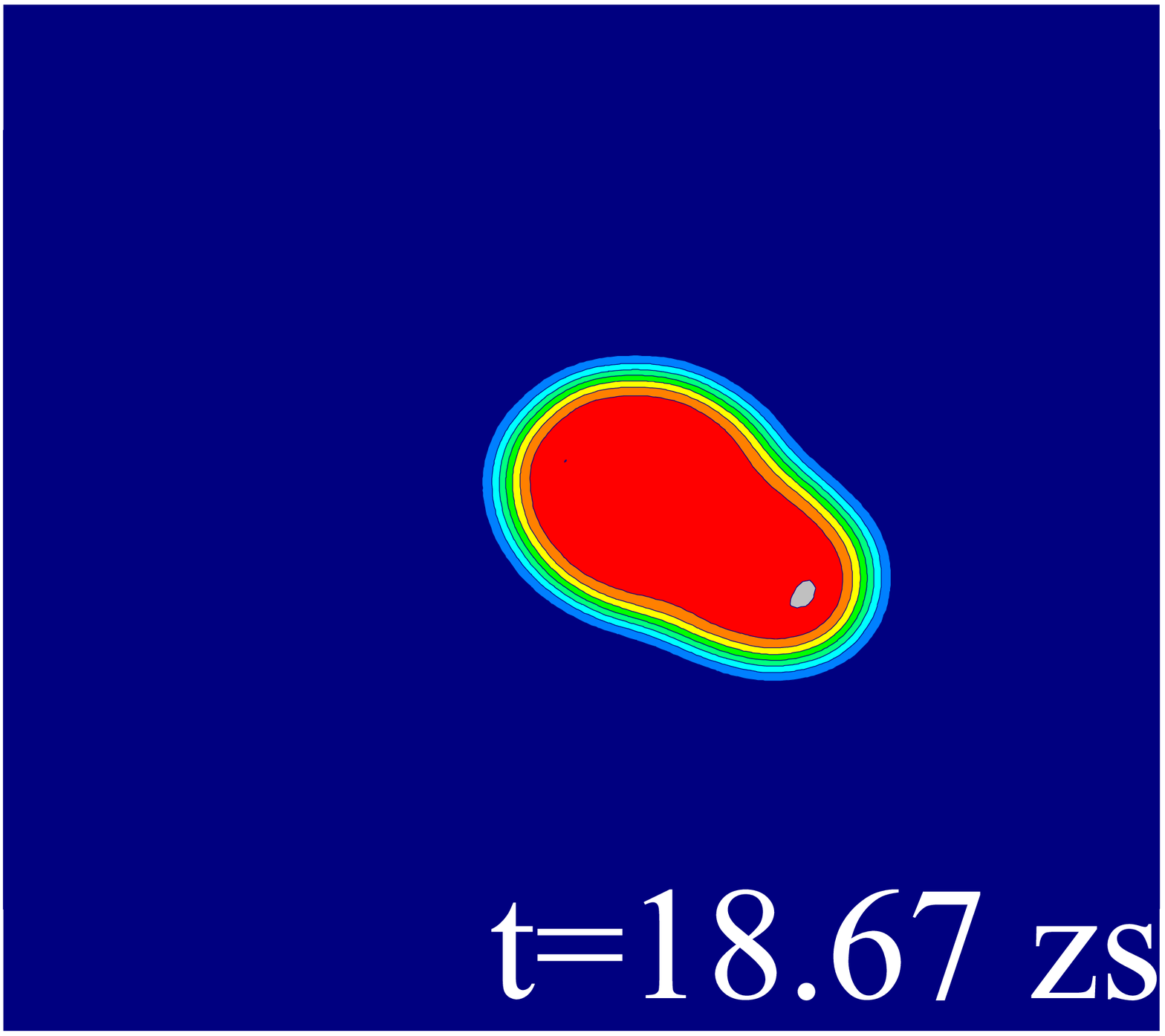}\hspace {1.2pt}
\includegraphics[scale=0.12]{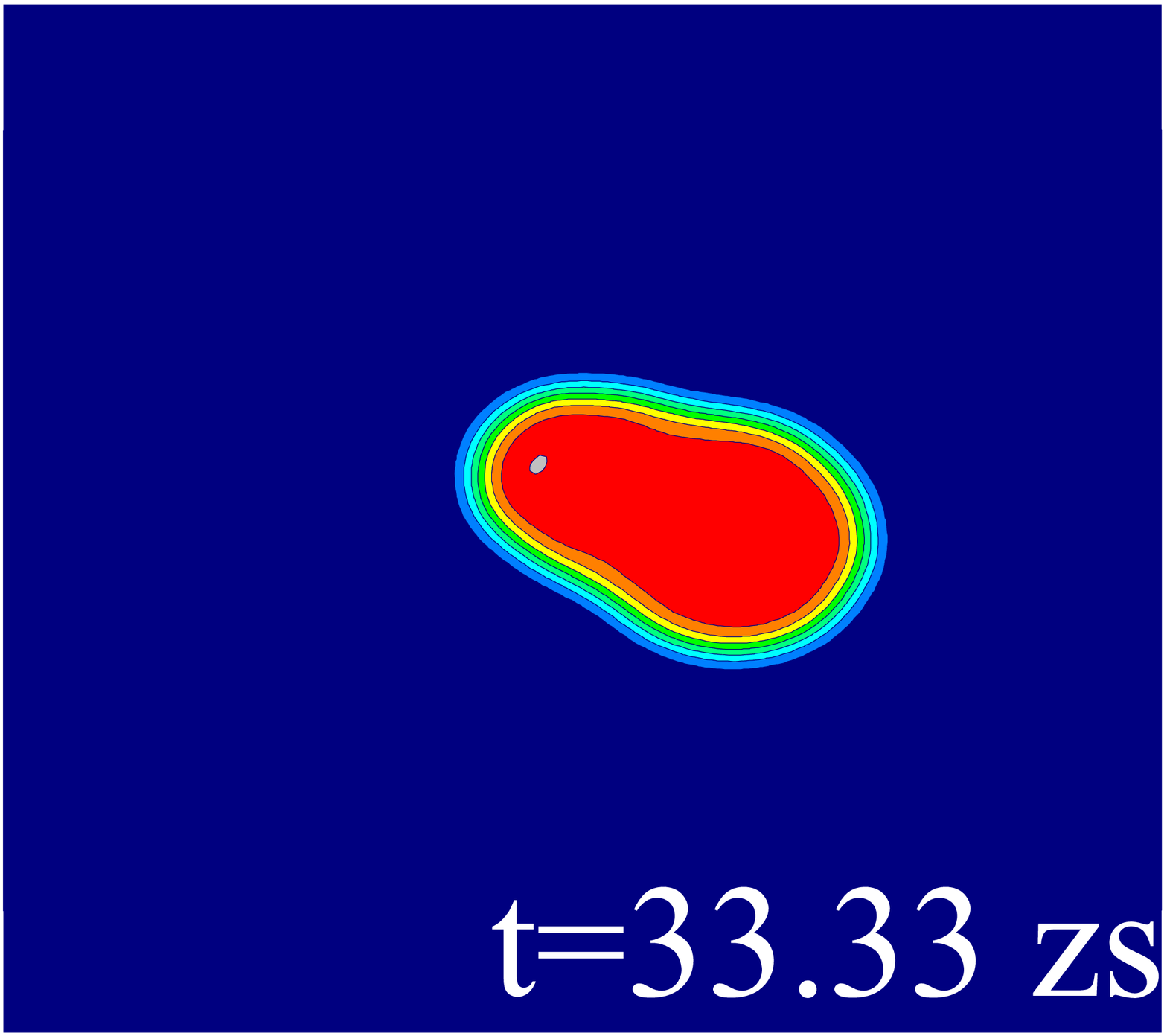}
\caption{(Color online) Time evolution of the mass density in the side collision of $^{48}
$Ca+$^{239}$Pu at $E_{\mathrm{c.m.}}=204.02$ MeV and $b$=1.5~fm.}
\label{fig:239Puside}
\end{figure}

We first consider the reaction $^{48}$Ca+$^{239}$Pu at $E_{\mathrm{lab}}$ =
245 MeV (corresponding to $E_{\mathrm{c.m.}}$ = 204.02 MeV), which is the
energy used in the Dubna experiment~\cite{Utyonkov2015_PRC92-034609}. The time
evolution of the mass density is displayed in Fig.~\ref{fig:239Puside}, in
which the symmetry axis of the prolate deformed nucleus $^{239}$Pu is
initially set perpendicular to the internuclear axis. This is the so-called
side orientation which leads to the largest contact time in central
collisions~\cite{Wakhle2014_PRL113-182502}. We observe that TDHF calculation
predicts fusion for this collision. Our definition for fusion
is that an event has the contact time larger than 35 zs, and furthermore a mononuclear
shape without any neck formation is required. As shown in Fig.~\ref{fig:239Puside}, in the early stage of collision,
there forms a neck between the two fragments, and the dinuclear system starts
to rotate. With the time evolution, more nucleons transfer from heavy to light
fragment and the neck grows into a mononuclear shape. This mononuclear system
keeps its shape for enough long time and results in the fusion process.

\begin{figure}[tb]
\centering
\includegraphics[scale=0.12]{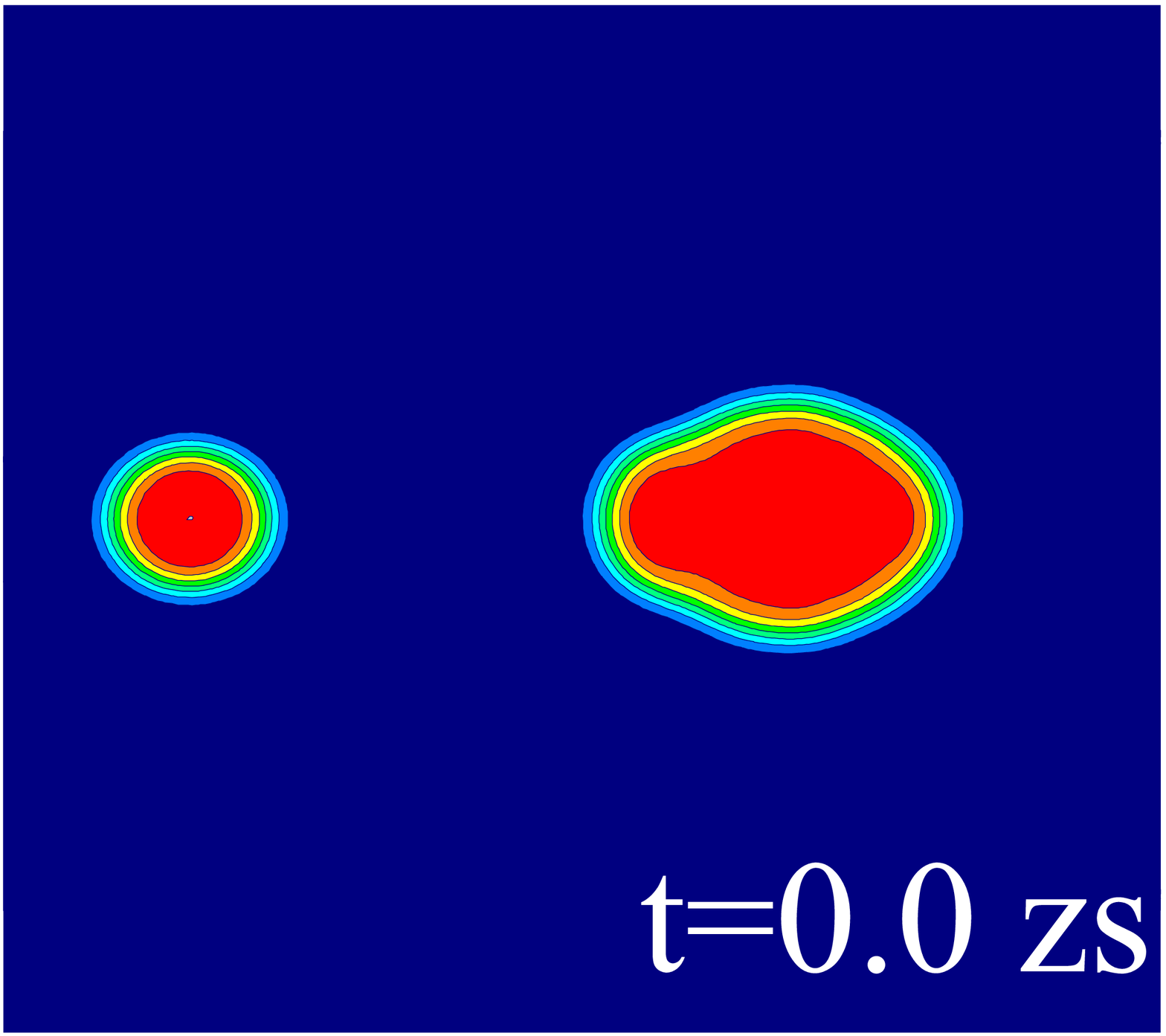}\hspace {1.2pt}
\includegraphics[scale=0.12]{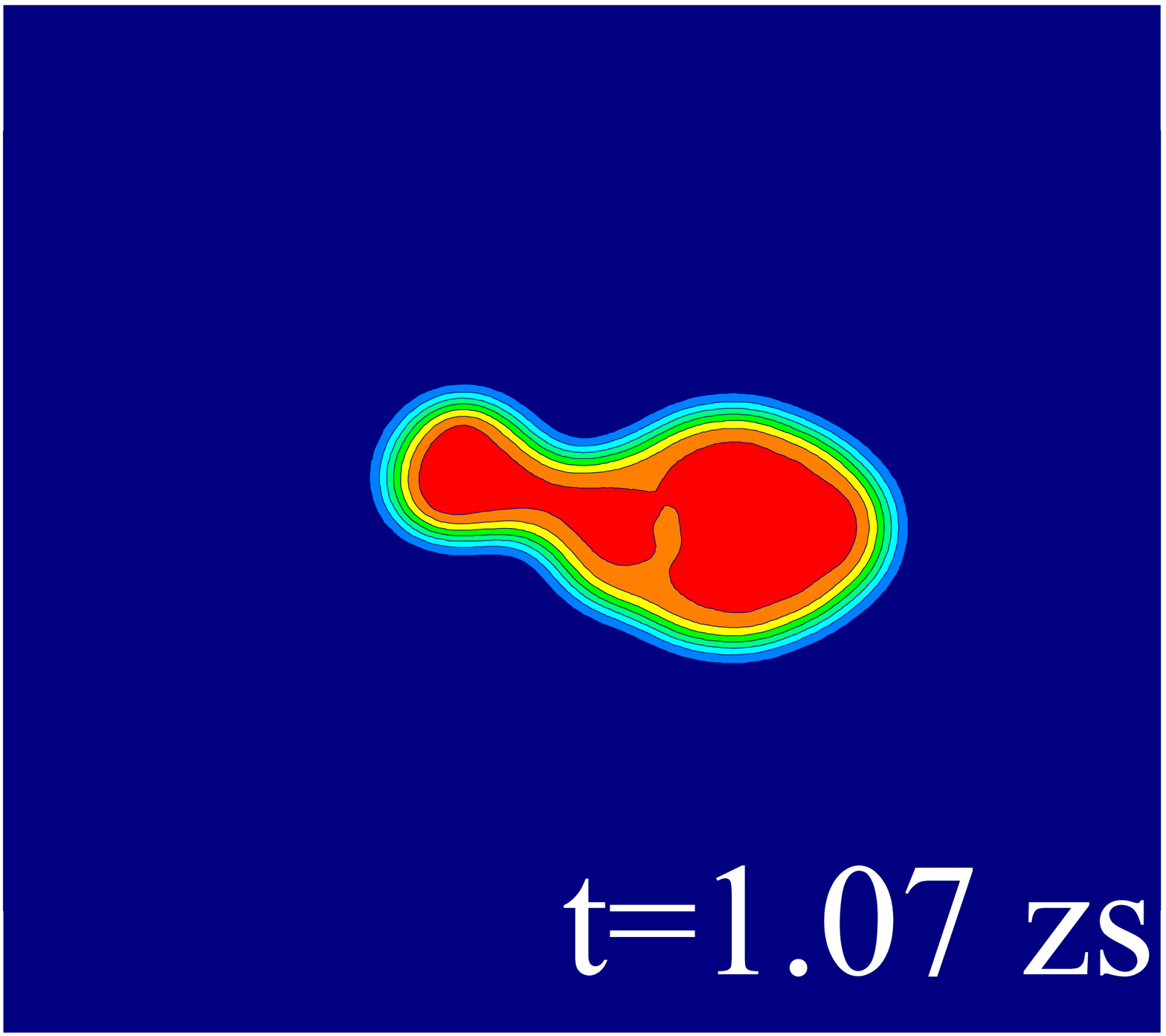}\hspace {1.2pt}
\includegraphics[scale=0.12]{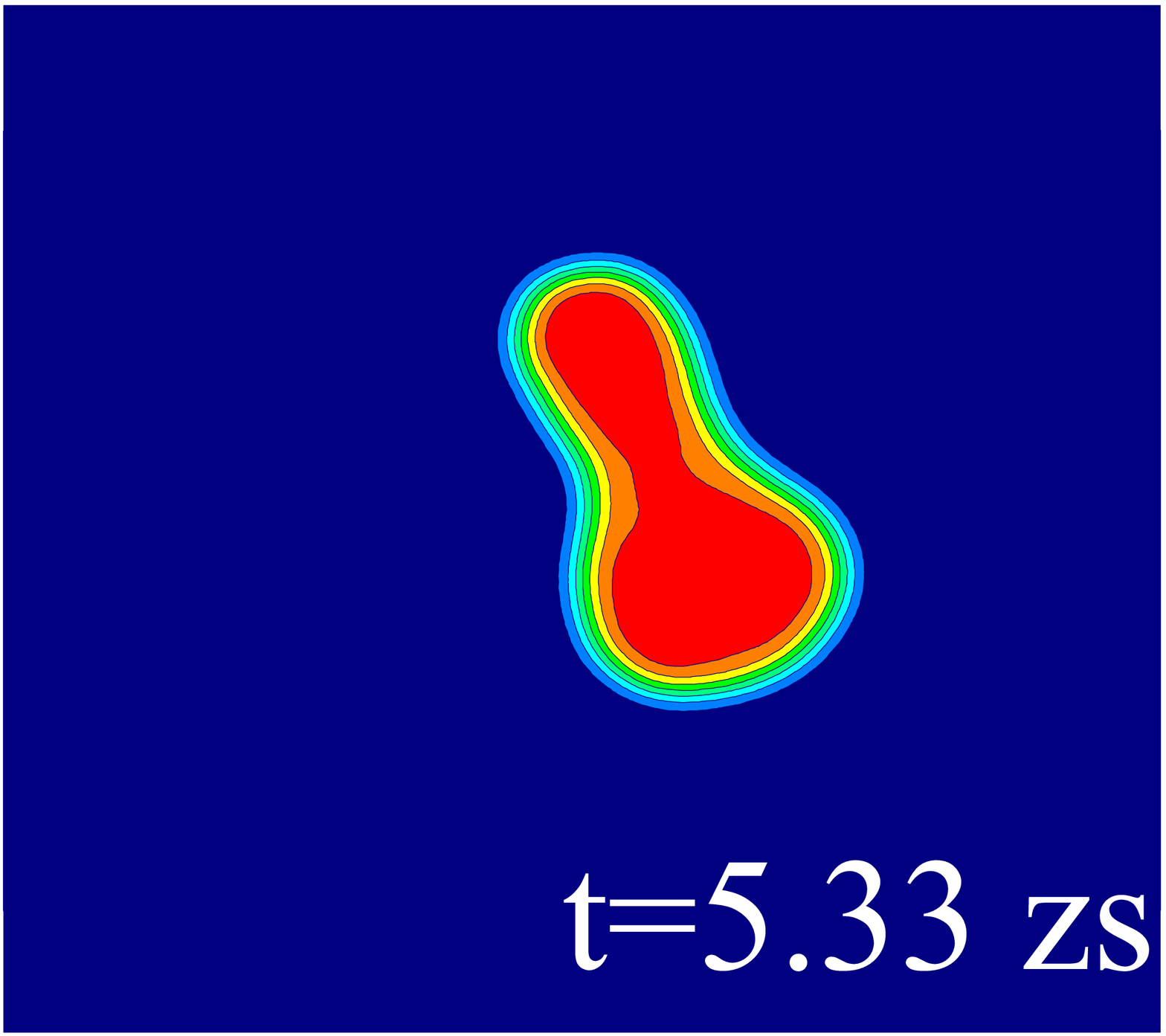}\\[-0.5pt]
\includegraphics[scale=0.12]{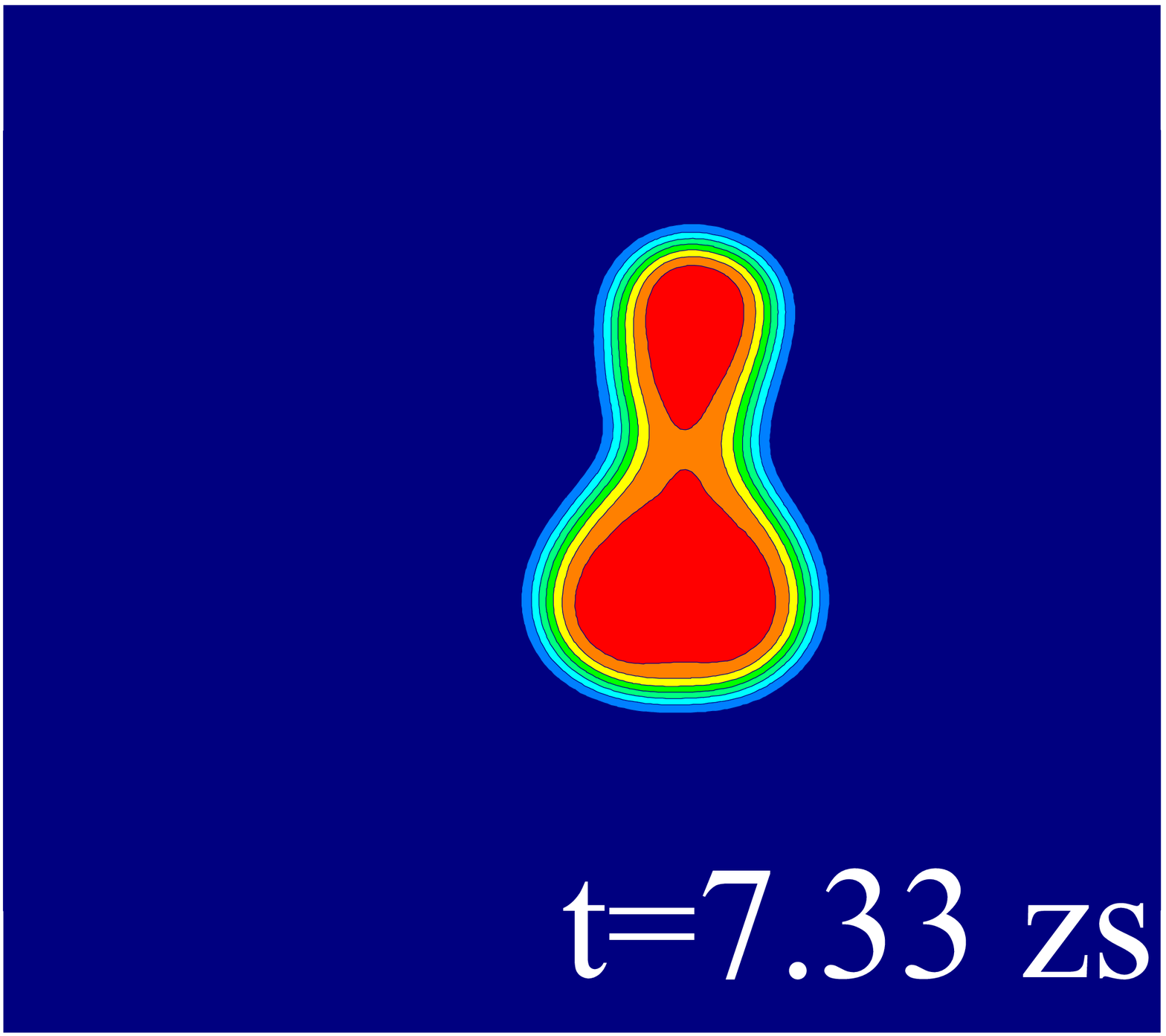}\hspace {1.2pt}
\includegraphics[scale=0.12]{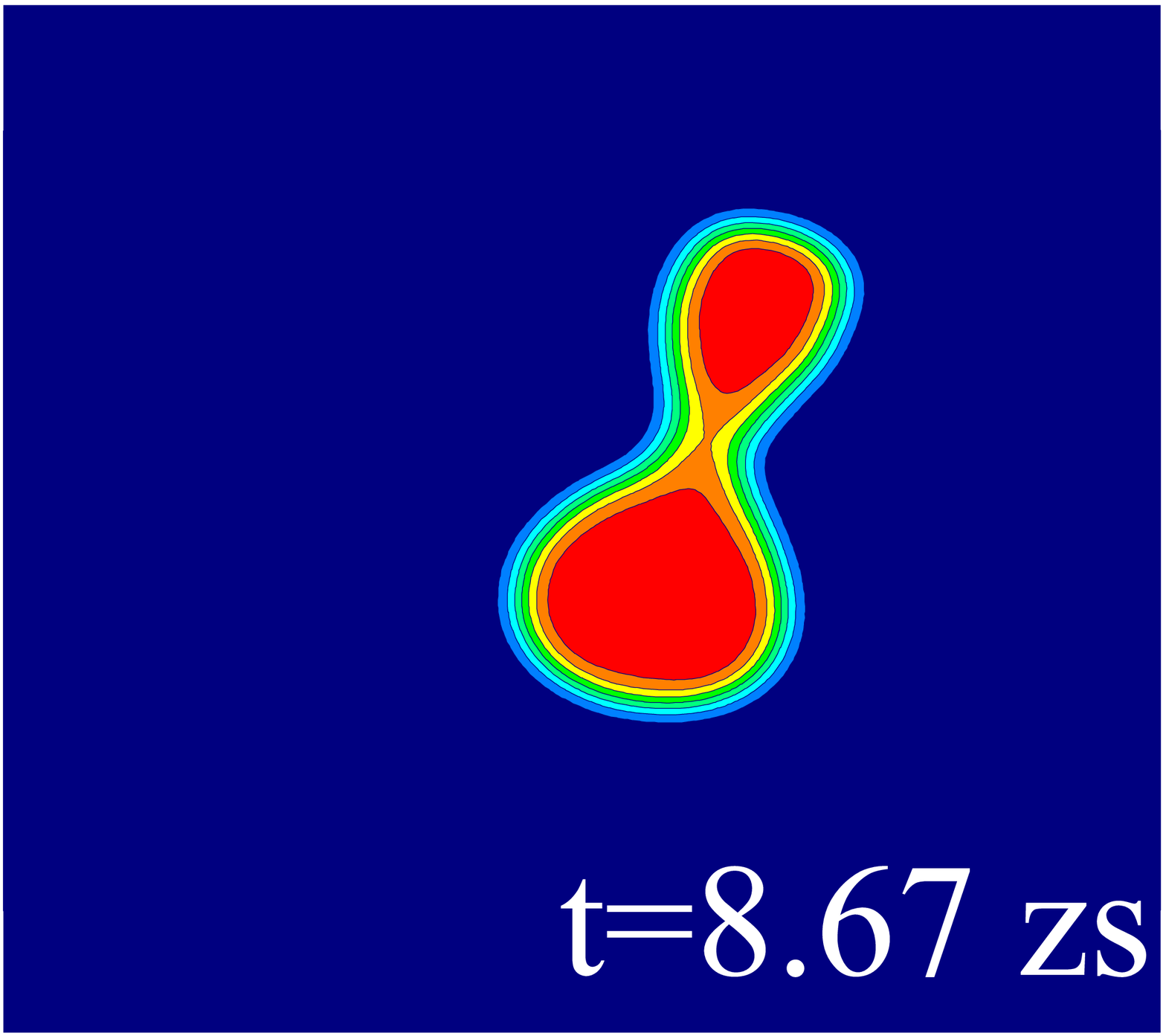}\hspace {1.2pt}
\includegraphics[scale=0.12]{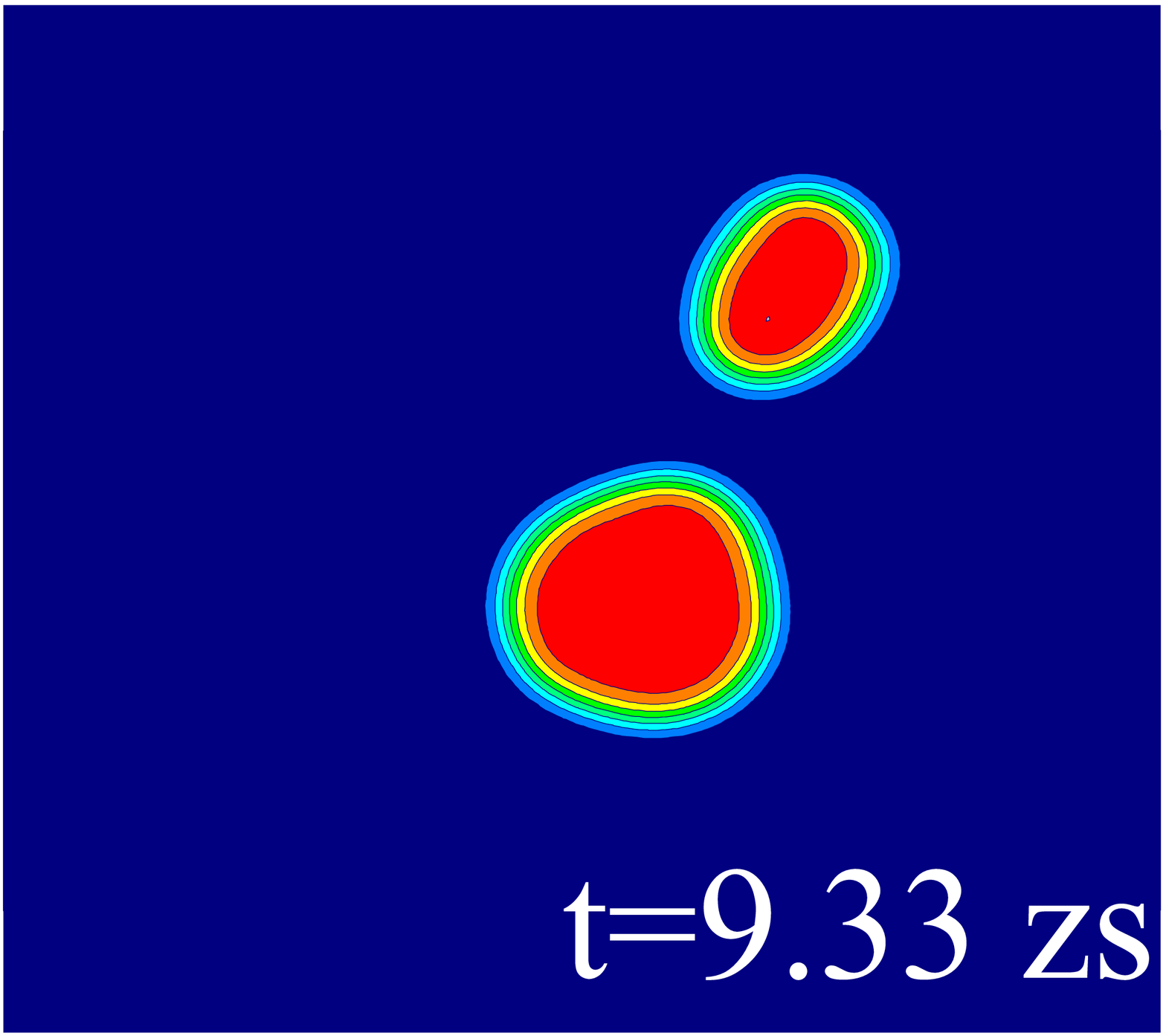}\caption{(Color
online) Time evolution of the mass density in the tip collision of
$^{48}$Ca+$^{239}$Pu at $E_{\mathrm{c.m.}}=204.02$ MeV and $b$=2.5~fm.}
\label{fig:239Putip}
\end{figure}

By contrast, quasifission is observed for the same reaction, but with the tip orientation and impact
parameter $b$=2.5~fm, as shown in Fig.~\ref{fig:239Putip}. Until the neck formation, the
density evolution is similar as in Fig.~\ref{fig:239Puside}. After that,
due to the competition between Coulomb repulsion and
centrifugal force, the dinuclear system prolongs and then breaks in two
fragments. This is the QF process which is characterized, as
compared to the deep inelastic collision (DIC), by the long contact time and
significant nucleon transfer. In this tip collision, the contact time is found to
be 8.17~zs and roughly 30 nucleons are transferred from $^{239}$Pu to $^{48}%
$Ca. Here the contact time is calculated as the time interval in which the
lowest density in the neck exceeds half of the nuclear saturation density
$\rho_{0}/2=0.08~\mathrm{{fm}^{-3}}$%
~\cite{Wakhle2014_PRL113-182502,Oberacker2014_PRC90-054605,Umar2016_PRC94-024605}%
. The light and heavy fragments in the exit channel are around $^{79}$Ge and
$^{208}$Pb, respectively. The heavy fragment lying around $N=82$ magic shell
indicates that the quantum shell effect plays an important role in the QF
dynamics. Such a multi-nucleon transfer is crucial to understand the
dissipation dynamics in heavy-ion collisions.

\begin{figure}[tb]
\includegraphics[width=7.8cm]{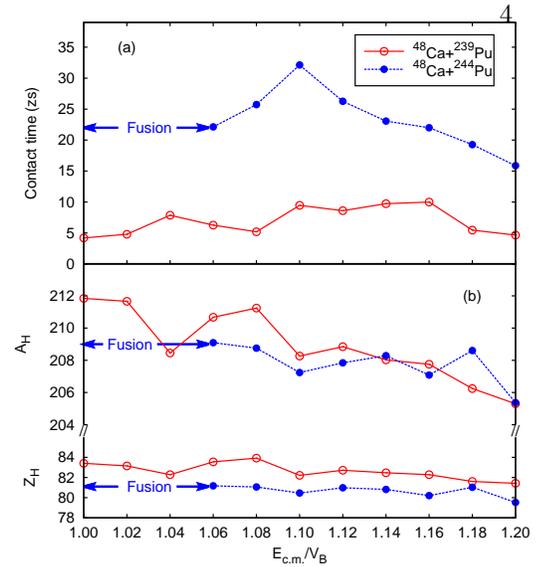}
\caption{(Color online) Contact time (a), mass and charge of the heavy fragments (b) as a
function of $E_{\mathrm{c.m.}}/V_{\mathrm{B}}$ for the central collisions
$^{48}$Ca+$^{239}$Pu (open circle) and $^{48}$Ca+$^{244}$Pu (solid circle)
with the tip of $^{239,244}$Pu.}
\label{fig:energy_dep_TAZ}
\end{figure}

We now compare the energy dependence of fusion and quasifission dynamics in the central collisions $^{48}$Ca+$^{239,244}$Pu.
The energy range is chosen as $E_{\mathrm{c.m.}}/V_{\mathrm{B}}=$1.0--1.2, denoted by the ratio
between the colliding energy $E_{\mathrm{c.m.}}$ and TDHF capture
barrier $V_{\mathrm{B}}$. The TDHF capture barrier is calculated for two
extreme orientations of the deformed targets $^{239,244}$Pu (tip and side). For
the tip collision, the barrier is found to be 177 MeV with $^{239}$Pu target
and 181 MeV for $^{244}$Pu, respectively. The side collision
results in a significantly higher barrier of 199 MeV both for $^{239,244}$Pu
targets, as expected. TDHF calculations show that both central collisions $^{48}$Ca+$^{239,244}$Pu
with the side of target nuclei lead to fusion within the above energy range.
Certainly, the noncentral collisions with the side orientation may show
quasifission. Hence, we leave out the results for the side collisions, and the
central collisions with the tip orientation are shown in
Fig.~\ref{fig:energy_dep_TAZ}. For the reaction $^{48}$Ca+$^{239}$Pu (open
circle), we observe that the contact time is within the value of 5--10 zs over
a wide range of energies. But, a dramatically different behavior appears for
$^{48}$Ca+$^{244}$Pu (solid circle). At energies close to barrier
($E_{\mathrm{c.m.}}/V_{\mathrm{B}}=1.0-1.06$.), the contact time is larger
than 35 zs and a mononuclear shape is kept, which is
considered as fusion reaction leading to the formation of compound nucleus.
As the energy increases, quasifission is observed, in which the contact time
is systematically larger than the $^{239}$Pu case.
The long contact time of the quasifission
process with the neutron-rich system relative to the neutron-deficient
system have also been observed in Cr$+$W reactions
~\cite{Hammerton2015_PRC91-041602}. In the dynamical evolution
of the neutron-rich system, the elongation of the dinuclear
system is much slower and the compact configuration with
mononuclear shape remains much longer, which are expected to lead
to a longer contact time. These results demonstrate
that the reaction with the neutron-rich target favors the fusion.
In similar reactions $^{48}$Ca+$^{238}$U~\cite{Oberacker2014_PRC90-054605} and $^{48}$Ca+$^{249}%
$Bk~\cite{Umar2016_PRC94-024605}, tip orientations are never found to lead to
fusion in TDHF calculations using SLy4d
force~\cite{Chabanat1998_NPA635-231,Chabanat1998_NPA643-441}. This difference
may arise from the choice of different Skyrme interaction,
numerical approximations, and the shell structure of target nuclei. In particular,
the influence of Skyrme interaction in QF should be examined. Since TDHF calculations for
reactions leading to superheavy systems require very long CPU times, the
interaction dependence of QF dynamics will be the subject of future works,
which is beyond the purpose of present study.

In the lower panel of Fig.~\ref{fig:energy_dep_TAZ} are the corresponding mass
and charge of the heavy fragments. We find that the transferred nucleon number
is within a small variation as a function of energy. For the reaction $^{48}
$Ca+$^{239}$Pu, the heavy QF fragments has the charge
$Z_{H}\simeq81.4-83.9$ and mass $A_{H}\simeq205.3-211.9$ in this energy
range, which is centered in the vicinity of the doubly magic nucleus $^{208}
$Pb. The tip collisions clearly favor the production of heavy fragments near
$^{208}$Pb at all energies, indicating a strong influence of the quantum shell
effects. A similar effect was observed in TDHF calculations for the tip
collisions using $^{238}$U~\cite{Oberacker2014_PRC90-054605} and $^{249}
$Bk~\cite{Umar2016_PRC94-024605} targets. For the reaction $^{48}
$Ca+$^{244}$Pu, the 1--2 more protons and 3--4 more neutrons are transferred
on average, as compared to $^{48}$Ca+$^{239}$Pu, due to the charge equilibrium.

\begin{figure}[tb]
\includegraphics[width=7.8 cm]{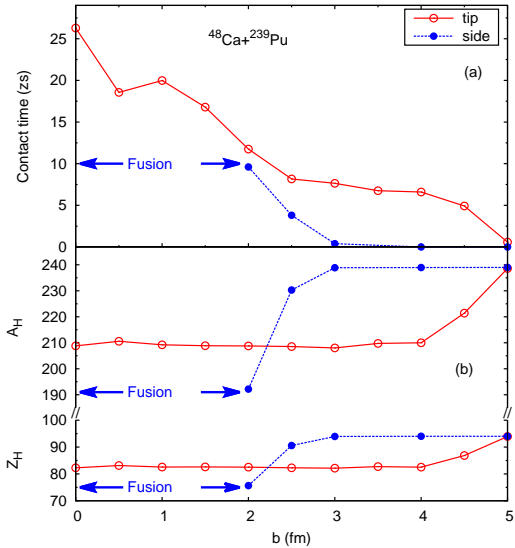}
\caption{(Color online) Contact time (a), mass and charge of the heavy fragments (b) as a
function of impact parameter both for the tip (open circle) and side (solid
circle) collisions of $^{48}$Ca+$^{239}$Pu at $E_{\mathrm{c.m.}}=204.02$ MeV used in Dubna experiment~\cite{Utyonkov2015_PRC92-034609}.}
\label{fig:ip_pu239TAZ}
\end{figure}

We next concentrate on the impact parameter dependence of fusion and quasifission
dynamics in connection with the experimental observation, in which the
isotopic dependence of measured ER cross section of Flerovium isotopes has been displayed in
Ref.~\cite{Utyonkov2015_PRC92-034609}. As stated in the Abstract of
Ref.~\cite{Utyonkov2015_PRC92-034609}, the measured ER cross section in $^{239}%
$Pu($^{48}$Ca,$3n$)$^{284}$FI reaction channel are about 50 times lower than
using $^{244}$Pu as target nucleus. However, the precise mechanisms of this
remarkable isotopic dependence are not well understood yet. In the following,
we will investigate how the QF and FF affect the production of SHEs.

The collision $^{48}$Ca+$^{239}$Pu at $E_{\mathrm{lab}}$ = 245 MeV
($E_{\mathrm{c.m.}}$ = 204.02 MeV), which is the energy used in the Dubna
experiment~\cite{Utyonkov2015_PRC92-034609}, is shown in
Fig.~\ref{fig:ip_pu239TAZ}. For the tip collision (open circle), the contact
time is roughly linear decreasing for the impact parameter
$b < 3$ fm, and then form a plateau with small
variance until $b$=4.0 fm. After that, a quick decrease of contact time is
observed which corresponds to the inelastic scattering. No fusion events are predicted for the tip
orientation. The side collision (solid circle) shows a very different
behavior as compared to the tip collision. We find that TDHF predicts fusion for
$b < 2$ fm, and then a rapid drop-off within a narrow range of impact
parameter 2\verb|-|3 fm. At $b$=3\verb|-|5 fm the quasielastic collisions happen,
characterized by the nearly zero contact time and identical fragments before
and after collision. In Fig.~\ref{fig:ip_pu239TAZ}(b) are the corresponding
mass and charge of the heavy fragments. For the tip collision, the transferred
nucleon is nearly constant in the quasifission region $b$=0\verb|-|4 fm,
producing the typically heavy fragments around $^{208}$Pb. This implies that the magic shell effect
is also observed to be crucial as a function of impact parameter
for the tip collisions. For the side collision, the transferred
nucleon is proportional to the contact time. We find that only
collision with the tip of $^{239}$Pu produces QF fragments in the magic $Z$=82
region, while collisions with the side are the only ones that may result in
fusion. These findings are consistent with the experimental observation in the
collision with $^{238}$U target~\cite{Wakhle2014_PRL113-182502}.

\begin{figure}[tb]
\includegraphics[width=7.8 cm]{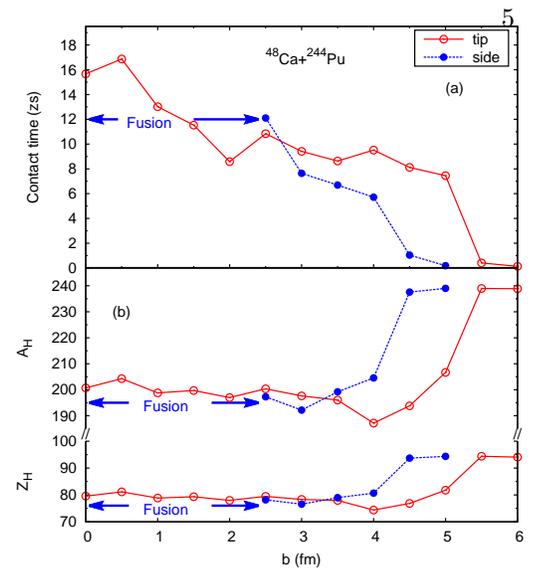}
\caption{(Color online) Contact time (a), mass and charge of the heavy fragments (b) as a
function of impact parameter both for the tip (open circle) and side (solid circle) collisions
of $^{48}$Ca+$^{244}$Pu at $E_{\mathrm{c.m.}}=216.76$ MeV used in GSI experiment~\cite{Gates2011_PRC83-054618}.}
\label{fig:ip_pu244TAZ}
\end{figure}

For comparison, the same observables for the reaction $^{48}$Ca+$^{244}$Pu at
energy $E_{\mathrm{lab}}$ = 259.4 MeV ($E_{\mathrm{c.m.}}$ = 216.76 MeV) used
in GSI-TASCA experiment~\cite{Gates2011_PRC83-054618} are displayed in
Fig.~\ref{fig:ip_pu244TAZ}. For the tip collision, the contact time behaves
similar as in $^{48}$Ca+$^{239}$Pu. The
multi-nucleon transfer from the heavy to light nucleus produces the heavy fragments with the charge
$Z_{H}\simeq74-82$ and mass $A_{H}\simeq192-212$. The heavy QF fragment is again in the vicinity of
Z=82 magic shell. For the side orientation of $^{244}$Pu nucleus, TDHF shows
fusion for $b<2.5$ fm and quasifission at $b$=2.5--4 fm, both of which are in a
larger angular momentum window as compared to the $^{239}$Pu case,
indicating a high possibility for fusion with $^{244}$Pu target.

By comparing the impact parameter dependence of fusion and quasifission
dynamics between the two reactions (Fig.~\ref{fig:ip_pu239TAZ} and
Fig.~\ref{fig:ip_pu244TAZ}), the fusion happens in the side collision for
$b<2.0$ fm with $^{239}$Pu target and $b<2.5$ fm with $^{244}$Pu. We further
refine the maximum impact parameter for fusion within the precision
of 0.1 fm, and find that the critical value is 1.7 fm and 2.3 fm for
$^{239}$Pu and $^{244}$Pu cases, respectively. The fusion window is about 0.6 fm
large in $^{48}$Ca+$^{244}$Pu, indicating that quasifission is
considerably reduced as compared to the $^{239}$Pu
case. By using the sharp cut-off formula Eq.~(\ref{FCS}), the ratio of fusion cross section
between $^{244}$Pu and $^{239}$Pu side collisions is roughly 1.8. One should note
that the average over the orientation for the deformed target nucleus should
be done if one wants to get an accurate value for this ratio.

\begin{figure}[tb]
\includegraphics[width=7.8 cm]{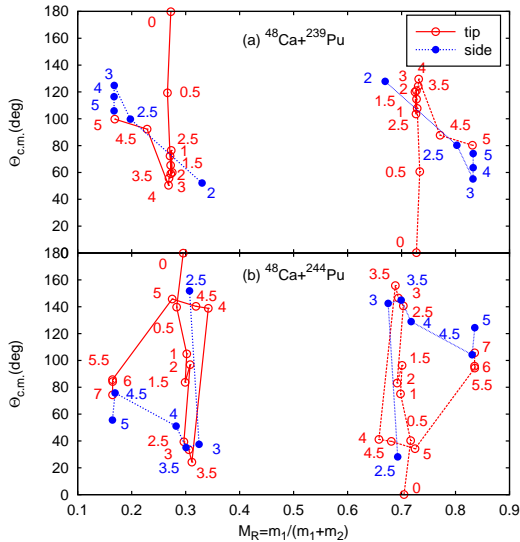}
\caption{(Color online) Mass-angle distribution of QF
fragments both for the tip (open circle) and side (solid
circle) collisions of $^{48}$Ca+$^{239}$Pu at $E_{\mathrm{c.m.}}=204.02$~MeV
(a) and $^{48}$Ca+$^{244}$Pu at $E_{\mathrm{c.m.}}=216.76$~MeV (b). The impact
parameters (in units of fm) are written next to each data point.}%
\label{fig:mass_angle}%
\end{figure}

On the other hand, the fusion-fission of compound nucleus dominated mainly by the height of
fission barrier and neutron evaporation may be sensitive to the neutron-richness of compound nucleus.
This FF process is at a much longer time-scale than QF and has no
memory of the entrance channel, which is beyond the scope of TDHF studies.
In order to calculate the ratio of ER cross sections between $^{292}$Fl ($4n$)
and $^{287}$Fl ($3n$), besides the ratio of their fusion cross section, we also need to
calculate the ratio of their survival probability $W_{\rm sur}$.
However during the decay, due to the very low fission barrier $B_{\rm f}$,
the fission dominates the decay process and has several orders higher probability
than the neutron emissions, and thus the $W_{\rm sur}$ after several neutron emissions
is several orders smaller than the fusion cross section~\cite{Shen2002_PRC66-061602}.
Empirically, 1 MeV larger of the $B_{\rm f}$
causes one order higher of the survival probability $W_{\rm sur}$.

Some theoretical investigations indicated that $^{298}$Fl could be the next double-magic nucleus
after $^{208}$Pb, and then one may expect more stable and larger $B_{\rm f}$
of $^{292}$Fl compared to $^{287}$Fl. This is confirmed by
several models predicting $B_{\rm f}$
\cite{Moller2009_PRC79-064304,Staszczak2013_PRC87-024320,Jachimowicz2017_PRC95-014303} and
giving the difference of the fission barrier
$\Delta B_{\rm f}$ (= $B_{\rm f}(^{292}$Fl) $- B_{\rm f}(^{287}$Fl))
around $0.82\sim 1.98$ MeV and averaging at $B_{\rm f}=1.2$ MeV.
Another simpler but frequently used $B_{\rm f}$
is approximated as $B_{\rm f(LD)} - E_{\rm{sh}}$, where $B_{\rm f(LD)}$ is the
liquid drop fission barrier and $E_{\rm sh}$ the shell correction energy.
As an example, by adopting the $B_{\rm f(LD)}$ from Ref. \cite{Dahlinger1982_NPA376-94}
and the $E_{\mathrm{sh}}$ from Ref. \cite{Wang2014_PLB734-215},
corresponding to $\Delta B_{\rm f}=0.55$ MeV, the $W_{\rm sur}(4n)$
for $^{292}$Fl at $E^{*} = 41.7$ MeV
is $8.21\times 10^{-8}$, while the $W_{\rm sur}(3n)$ for $^{287}$Fl
at $E^{*}=37.7$ MeV is $5.5\times 10^{-9}$,
and hence the ratio of the survival probability for the two reactions is 14.9. Finally,
involving estimated ratio of fusion cross sections,
the approximate ratio of ER cross sections between the two reactions is
roughly 27, close to the ratio the experimental data showing.

\begin{figure}[tb]
\includegraphics[width=7.8 cm]{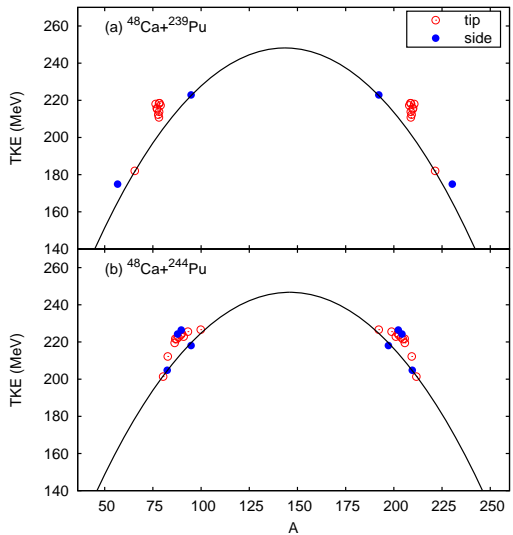}
\caption{(Color online) TKE-mass distribution of QF fragments both
for the tip (open circle) and side (solid
circle) collisions of $^{48}$Ca+$^{239}$Pu at $E_{\mathrm{c.m.}}=204.02$~MeV
(a) and $^{48}$Ca+$^{244}$Pu at $E_{\mathrm{c.m.}}=216.76$~MeV (b). The Viola
systematics (solid line) is also shown.}
\label{fig:mass_kinetic}
\end{figure}

The mass-angle distributions (MADs) of QF fragments have been measured in
experiments, which can be used to distinguish the QF from FF events.
In experiments, FF fragments are usually more symmetric than in QF. In
Fig.~\ref{fig:mass_angle}, we show the scattering angle as a function of the
mass ratio $M_{\mathrm{R}}=m_{1}/(m_{1}+m_{2})$, where $m_{1}$ and $m_{2}$ are
the masses of QF fragments, both for the tip (open circle) and side (solid
circle) collisions of $^{48}$Ca+$^{239,244}$Pu. The values of impact
parameters are written next to each data point. The scattering angle is
calculated as the sum of the incoming and outgoing Coulomb scattering angles
plus the TDHF scattering angle in the center-of-mass frame. We observe that
for the tip collisions of $^{48}$Ca+$^{239}$Pu the mass ratio of QF events is
near $M_{\mathrm{R}}=$0.22--0.27 for light fragments and $M_{\mathrm{R}}%
=$0.73--0.78 for heavy fragments. At larger impact parameter $b$=4.5\verb|-|5~fm for
the tip collisions and $b$=2.5\verb|-|5~fm for the side collision, the light
fragments lying around $M_{\mathrm{R}}=$0.16 and heavy fragments $M_{\mathrm{R}}%
=$0.84 correspond to the quasielastic and deep-inelastic reactions.
One should note that TDHF is a deterministic theory which gives only the most probable event
instead of a probability distribution of various events.
In the lower panel of Fig.~\ref{fig:mass_angle},
the MADs in $^{48}$Ca+$^{244}$Pu show similar behavior in comparison to the
$^{48}$Ca+$^{239}$Pu case, except the larger mass ratio due to the more
nucleon transfer. The MADs for the side orientation span a larger angular
range in $^{48}$Ca+$^{244}$Pu. The detailed differences can be
recognized by the values of the MADs for the specific impact parameters.

The correlation between mass and total kinetic energy (TKE) of QF fragments
is another key experimental observable to distinguish the quasielastic events
from fully damped events such as QF and FF. In TDHF, the
TKE is calculated as the sum of kinetic energy of the fragments after the
separation and the Coulomb potential energy assuming that the fragments are
pointlike charges. Figure~\ref{fig:mass_kinetic} shows that the TKE-mass
distribution of QF fragments both for the tip (open circle) and side (solid
circle) collisions of $^{48}$Ca+$^{239,244}$Pu. The Viola systematics (solid
line)~\cite{Viola1985_PRC31-1550,Hinde1987_NPA472-318} is also shown and has
been known in good agreement with the measured TKE of experimental fission
fragments. We observe that the TKE of QF fragments are distributed around the
Viola systematics, indicating that most of the relative kinetic energy has
been dissipated in the collision.

\section{Summary}

\label{summary}

We investigate the isotopic dependence of quasifission and fusion-fission in
connection with the experimental production of superheavy flerovium isotopes
for the reactions $^{48}$Ca+$^{239,244}$Pu by using the microscopic TDHF
approach and the statistical evaporation model HIVAP.
Quasifission dynamics may be characterized by many variables, for instance,
nuclear contact times, masses and charges as well as kinetic energies of the
fragments, and the mass-angle distribution. We investigate the dependence of
these variables on bombarding energy, impact parameter,
and orientation of the deformed target.
Long contact times associated with fusion are observed
in both reactions with a side orientated target nucleus,
whereas only quasifission dynamics is observed for the tip orientation. For
the tip collisions, the quasifission fragments lying in the vicinity of the
magic nucleus $^{208}$Pb indicate the importance of the quantum shell effect.
No quantum shell effects are observed in the collisions with the side of
$^{239,244}$Pu targets. The calculated mass-angle and TKE-mass distributions of QF
fragments can be used for a direct comparison with the experimental
measurements.

The quasifission in the reaction $^{48}$Ca+$^{244}$Pu is found to be
remarkably reduced as compared to the $^{48}$Ca+$^{239}$Pu
case, which leads to the ratio of fusion cross section between $^{244}$Pu and $^{239}$Pu side collisions roughly 1.8.
The differences are attributed to the neutron-richness of target nucleus.
While considering the isospin-dependence of the fission barrier, the ratio of
survival probability of $^{292}$Fl ($4n$) to $^{287}$Fl ($3n$)
is about 14.9, and thus the predicted ratio of their residue cross sections is about 27,
close to the value shown by experimental data.
Therefore, our studies explain the experimental observations and make clear of
the precise reaction mechanisms, which may motivate the
experimentalists to produce SHEs with more efficient target-projectile combinations.
The present method TDHF+HIVAP demonstrates its usefulness for the microscopic reaction mechanism
of quasifission and fusion-fission dynamics.

\section{Acknowledgments}

L. G. thanks A. S. Umar for helpful discussions. This work is
partly supported by NSF of China (Grants No. 11175252, 11575189, 11747312,
U1732138, 11790325, 11790323), NSFC-JSPS International Cooperation Program (Grant No.
11711540016), and Presidential Fund of UCAS. The computations in present work
have been performed on the High-performance Computing Clusters of
SKLTP/ITP-CAS, Tianhe-1A supercomputer located in the Chinese National
Supercomputer Center in Tianjin, and the C3S2 computing center in Huzhou University.

\bibliographystyle{apsrev4-1}
\bibliography{ref}
\end{document}